\documentclass[aps,prl,reprint,groupedaddress]{revtex4-1}
\usepackage[T1]{fontenc}
\usepackage[utf8]{inputenc}
\usepackage{graphicx}
\usepackage{amsmath}
\usepackage{amssymb}
\usepackage{color} 
\usepackage{pifont}
\usepackage{natbib}

\begin{document}
\title{Intra- and Intermolecular Inhomogeneities in the Course of Glassy Densification of Polyalcohols}

\author{Jan Philipp Gabriel \footnote{present}}
%\email{Jan.Gabriel@gmx.de}
\thanks{These authors contributed equally.}
\affiliation{Peter Debye Institute for Soft Matter Research, Leipzig University, 04103 Leipzig, Germany}

\author{Martin Tress \normalfont{\textsuperscript{*}}}
%\thanks{These authors contributed equally.}
\email[Correspondence to: ]{martin.tress@uni-leipzig.de}

\affiliation{Peter Debye Institute for Soft Matter Research, Leipzig University, 04103 Leipzig, Germany}

\author{Wilhelm Kossack}
\affiliation{Peter Debye Institute for Soft Matter Research, Leipzig University, 04103 Leipzig, Germany}

\author{Ludwig Popp}
\altaffiliation[Present address: ]{TUD}
\affiliation{Peter Debye Institute for Soft Matter Research, Leipzig University, 04103 Leipzig, Germany}

\author{Friedrich Kremer}
\email[Correspondence to: ]{fkremer@physik.uni-leipzig.de}
\affiliation{Peter Debye Institute for Soft Matter Research, Leipzig University, 04103 Leipzig, Germany}

\date{\today} 

\begin{abstract}
To test basic assumptions inherent to most theories of molecular liquids and glasses, Infrared spectroscopy is carried out on short polyalcohols at temperatures ranging from far above to far below their glass transition. By analyzing specific vibrations, the thermal expansion of covalent bonds and hydrogen bridges is determined revealing striking differences. A comparison with density verifies the negligibility of \textit{intra}molecular expansion but exposes severe inhomogeneities on \textit{inter}molecular scale. This signals distinct roles in glassy densification and promotes an atomistic understanding.
\end{abstract}

\maketitle 

%%%%%%%%%%%%%%%%%%%%%%%%%%%%%%%%%%%%%%
%Introduction
%%%%%%%%%%%%%%%%%%%%%%%%%%%%%%%%%%%%%%
% Limit for PRL: < 3750 words

The vitrification of a (super-cooled) liquid into a disordered solid is a central subject of experimental and theoretical soft condensed matter research  \cite{Wong1976,Anderson1979,Donth1981,Zallen1983,Elliott1990,Donth1992,Donth2001,Kremer:2002a,Lubchenko2007,Ngai2011,Gotze2012,Kremer:2018a}. A distinctive feature of super-cooled liquids and glasses is their diverging molecular fluctuation rate. Even simple concepts like the free volume model \cite{Fox1950,Fox1951,Fox1954,Turnbull1961} relate this to the reduction of intermolecular volume, in other words an increase in density, reflected in the term glassy densification. While the free volume model uses the inhomogeneity of the density only conceptually, early microscopic theories of liquids identify it as an additional key quantity \cite{Bernal1937}. This has also been established for glasses \cite{Ediger1998,Donth2002,Rissanou2015}, and it became clear that heterogeneity in time and space, in other words a varying temporal and spatial scale on which molecules fluctuate \cite{Ediger:2000a,Richert:2002a} is a fundamental attribute of both liquids and glasses. Consequently, modern theoretical approaches like the random first order transition (RFOT) theory \cite{Xia2000,Stevenson2005,Lubchenko2007} incorporate an inhomogeneous density in their description.

Interestingly, even RFOT theory, one of the most sophisticated concepts of molecular glasses, is based on abstract beads and their packing density \cite{Lubchenko2007}, i.e. particular intermolecular interactions are not specifically considered. Hence, this approach assumes that the impact of intermolecular interactions on structural relaxation is described by (packing) density. This means that certain presumptions are inherent to this model (similar to most other theoretical descriptions \cite{Dyre2006,Mirigian2014,Mauro2009,Lubchenko2007,Bosse1978,Adam1965}): i) \textit{intra}molecular expansion is neglected \cite{footnote_exp-neg}, i.e. density changes are fully or almost fully ascribed to \textit{inter}molecular expansion, and ii) no directionality or inhomogeneity of the \textit{inter}molecular interactions is accounted for, i.e. the thermal expansion even on molecular scale is considered isotropic. While the first conjecture appears rather plausible, the second one is questionable for a vast range of systems involving directional interactions e.g. hydrogen (H-) bridges as in glycerol, one of the most studied model glass formers \cite{Schneider:1998,Wuttke:1994,Klieber:2013,Jensen:2018,Capponi2010,Ryabov:2003,Beevers:1980,Yee:1996,gabriel2019a}. Despite their essential nature, to the best of our knowledge these two conjectures, negligibility of \textit{intra}molecular expansion and \textit{inter}molecular bond directionality, have not yet been tested experimentally.

Here, we reexamine our previously published Infrared (IR) spectroscopy data of short polyalcohols from a wide temperature range above and below the calorimetric glass transition temperature $T_g$ \cite{kremer2018glassy}. These molecules (glycerol, threitol, xylitol, and sorbitol) have a regular structure but the orientation of the hydroxy groups, which associate and form a H-bridge network, gives rise to a significant anisotropy. This \textit{intra}molecular inhomogeneity leads to weak and strong \textit{inter}molecular H-bridges, and with a novel extensive analysis we reveal their individual thermal expansions.  Despite the supramolecular network formed by the mildly expanding strong H-bridges, the profoundly expanding weak H-bridges surprisingly allow for considerable changes in density. A consequence of this inhomogeneity of the \textit{inter}molecular interactions is a decoupling of the structural relaxation (governed by strong bridges) from the thermal expansivity and density (dominated by weak bridges).

\begin{figure}
\centering
\includegraphics[width=8cm]{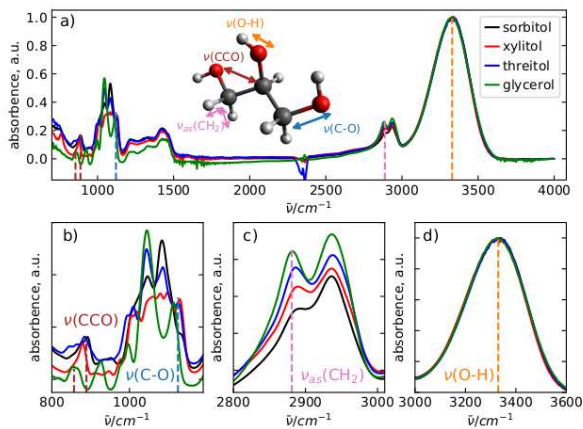}
\caption{a) IR absorbance spectra of several polyalcohols as indicated at selected temperatures (sorbitol: 249.4\,K, xylitol: 252.5\,K, threitol: 269.2\,K, and glycerol: 269.2\,K) to show identical spectral positions of the $\nu$(OH) band (dashed orange line); the absorbance is normalized to the $\nu$(OH) peak. Close-ups of the spectral regions of b) the $\nu$(CCO) band (brown dashed line) and the $\nu$(CO) band (blue dashed line), c) the $\nu_{as}$(CH$_2$) band (pink dashed line), and d) the $\nu$(OH) band (orange dashed line). The colored arrows in the molecular model of glycerol in a) exemplify the respective vibrations. The experimental uncertainty is smaller than the line width.}
\label{fig1}
\end{figure}

IR spectra were recorded using a Fourier transform infrared (FTIR) spectrometer (Bio–Rad FTS 6000) combined with an IR microscope (UMA 500) and a liquid nitrogen-cooled mercury-cadmiumtelluride (MCT) detector (Kolmar Technologies, Inc, USA) while a THMS 350V stage (Linkam Scientific Instruments, UK) flushed with dry nitrogen controlled the sample temperature; further information can be found elsewhere \cite{kremer2018glassy}. Comparing the spectra of the polyalcohols under study reveals their molecular characteristics (Fig.\,\ref{fig1}a). In the low frequency range ($\bar\nu=800-1200\,cm^{-1}$) the CO stretching vibration $\nu$(CO) at $\bar\nu\approx1125\,cm^{-1}$ and a combined CCO stretching vibration $\nu$(CCO) at $\bar\nu\approx860-890\,cm^{-1}$ exhibit bands distinctive for each substance (Fig.\,\ref{fig1}b). The $\nu$(CO) vibration shows only minor changes, i.e. a shift to higher frequencies with increasing molecular weight by $\sim20\,cm^{-1}$ from glycerol to sorbitol. In contrast, the $\nu$(CCO) band not only shifts by $\sim40\,cm^{-1}$ but also undergoes pronounced changes in shape for each material due to superimposing complex structural oscillations. Density functional theory (DFT) calculations of glycerol reveal distinct spectra for the various conformers in this frequency range \cite{Chelli2000}. Since a change in  molecular weight alters not only the number of conformers but also their probability distribution, such spectral differences are expected, and, due to their complexity, the absorption bands cannot be unraveled any further.

At higher frequencies, the vibrations become more localized and hence less complex, but also less distinct in the homologous series. Thus, the spectral region indicative of CH stretching vibrations ($\bar\nu=2800 -3000\,cm^{-1}$) exhibits only minor differences (Fig.\,\ref{fig1}c). The asymmetric CH$_2$ stretching vibration $\nu_{as}$(CH$_2$) at $\bar\nu\approx2880\,cm^{-1}$ shifts by less than $5\,cm^{-1}$ from glycerol to sorbitol; its absorbance reduces with increasing molecular weight since CH$_2$ stretching happens only at terminal carbons. Conversely, the absorbance of CH stretching $\nu$(CH) (from non-terminal carbons) at $\bar\nu\approx2950\,cm^{-1}$ increases; however, since it is superimposed by the symmetric CH$_2$ stretching vibration $\nu_{sm}$(CH$_2$) at $\bar \nu\approx2940\,cm^{-1}$ and the tail of the OH stretching vibration band $\nu$(OH), only $\nu_{as}$(CH$_2$) is analyzed quantitatively. At even higher frequencies ($\bar \nu=3000-3600\,cm^{-1}$), the broad $\nu$(OH) peak, typical for alcohols \cite{Bauer2015a}, at $\bar\nu\approx3350\,cm^{-1}$ is prominent and identical in shape in all four polyalcohols (Fig. \ref{fig1}d).

\begin{figure}
\centering
\includegraphics[width=8cm]{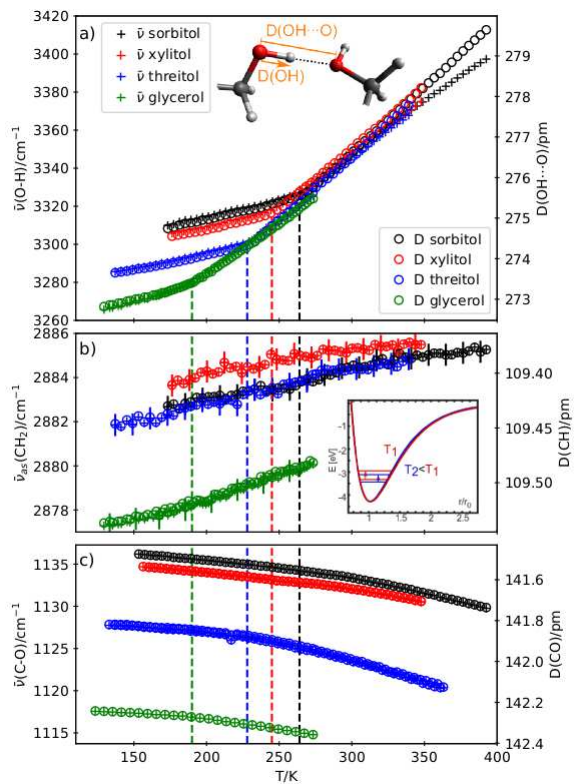}
\caption{Temperature dependence of the IR absorption frequencies of the a) $\nu$(OH), b) $\nu_{as}$(CH$_2$), and c) $\nu$(CO) bands (crosses, left axis) for different polyalcohols as indicated. The conversion of stretching frequencies into lengths (circles, right axis) of hydrogen bridges and covalent bonds is done via eq.\,(\ref{ohofunction}) for $\nu$(OH), and via a calibrated eq.\,(\ref{eq_nu_a_D}) for $\nu$(CH) and $\nu$(CO) (see text), respectively. Dashed vertical lines indicate each material's $T_g$ (same color code). The inset sketches vibrational states in the Morse potential at two different temperatures. The experimental uncertainty is smaller than the symbol size unless indicated otherwise.}
\label{fig2}
\end{figure}

From the spectra, peak frequencies of the $\nu$(OH), $\nu_{as}$(CH$_2$), and $\nu$(CO) band were determined in a wide temperature range (Fig.\,\ref{fig2}). The $\nu$(OH) band exhibits a large red-shift with decreasing temperature (Fig.\,\ref{fig2}a) amounting to $\sim60\,cm^{-1}/100\,K$ above $T_g$ in all substances. Further, the temperature dependence of the peak frequency has a distinct change of slope at $T_g$ (similar to density). In contrast, the red-shift upon temperature reduction observed in the $\nu_{as}$(CH$_2$) band is as small as $3\,cm^{-1}$ in the entire range ($150-250\,K$), and shows no distinct feature at $T_g$ (Fig.\,\ref{fig2}b). The $\nu$(CO) bands exhibit a small ($\sim5\,cm^{-1}$ in the whole range) blue-shift as temperature decreases (Fig.\,\ref{fig2}c) with a gradually changing slope above $T_g$ (though not as distinct as in $\nu$(OH)).

A shift in IR absorption frequency of a stretching vibration typically reflects a change in bond length \cite{footnote_IR,Schrader1995}. Consequently, the large shift in $\nu$(OH) means that the overall \textit{intra}molecular thermal expansion is dominated by changes in OH length while variations in CH and CO lengths have only minor impact. Since a red-shift indicates an expansion (weakening of the bond), both OH and CH bond lengths increase upon cooling (Fig.\,\ref{fig2}a \& b). This is caused by the respective \textit{inter}molecular interactions, i.e. strong (OH$\cdots$O) and weak (CH$\cdots$O) H-bridges. A densification with decreasing temperature reduces \textit{inter}molecular distances and brings the oxygen atoms of surrounding molecules (H-bridge acceptors) closer. Hence, the (\textit{inter}molecular) H-bridges become stronger while the (\textit{intra}molecular) covalent bonds are stretched \cite{footnote_neutrons,*Towey:2011}. In contrast, the blue-shift upon cooling in $\nu$(CO) (Fig.\,\ref{fig2}c) implies a contraction of the covalent bonds in the backbone (i.e. normal thermal expansion).

In a simple quantitative approach, the change in bond length is approximated by employing the Morse potential \cite{Morse:1929,Dahl:1988,Lima:2005} to describe molecular vibrations (Fig.\,\ref{fig2} inset):
\begin{equation}
 V(r) = E_d \left( 1- \exp\left[ -a (r-r_0)\right] \right)^2
\label{eq_morsepot}
 \end{equation}
Here, $E_d$ represents the dissociation energy and $a$ the inverse width of the potential. The latter can be related analytically to the difference of the energy eigenvalues in the ground state and the first excited state $\Delta E(\bar\nu)$:
\begin{equation}
a(E(\bar\nu),E_d) = \sqrt{\frac{\mu}{2\hbar^2}} \left( \sqrt{E_d} - \sqrt{E_d - 2\Delta E(\bar\nu)} \right)
 \label{eq_nu_a_D} 
 \end{equation}

Therein, $\mu$ is the reduced mass and $\hbar$ the reduced Planck constant. Using typical dissociation energies (e.g. from dimethylether $E_d(CO)=325\,kJ/mol$ \cite{Pacey1974} and methane $E_d(CH)=435\,kJ/mol$ \cite{Benson1970}) and inserting the measured IR absorption frequencies $\bar\nu(T)$ via $\Delta E(\bar\nu(T)) = 2\pi c\hbar \bar\nu(T)$ yields $1/a(T)$, an estimate for the change in interatomic (i.e. \textit{intra}molecular) distance with temperature. In order to approximate the absolute length of a bond $D(T)$ (Fig.\,\ref{fig2}b \& c), the respective interatomic distance obtained from DFT calculations \cite{orca2} of glycerol $D^{gly}$ were used for a calibration $D^{gly}=D(T_{ref})=1/a(T_{ref})+D_0$ at the reference temperature $T_{ref}=300\,K$ with the correction $D_0$ ($D_{CH}=109.47\,pm$, and $D_{CO}^{gly}=142.4\,pm$ with $D_{0,CH}=54.8\,pm$, and $D_{0,CO}^{gly}=96.4\,pm$ respectively). Extended DFT calculations to reproduce a temperature dependence fail since \textit{inter}molecular interactions are missing. (The latter are required to model an equilibrated, i.e. energetically minimized bond at different lengths). Therefore, eq.\,(\ref{eq_nu_a_D}) is used to quantify the bond lengths $D_{CH}$ and $D_{CO}$ \cite{Steiner1995}.

\begin{figure}
\centering
\includegraphics[width=8cm]{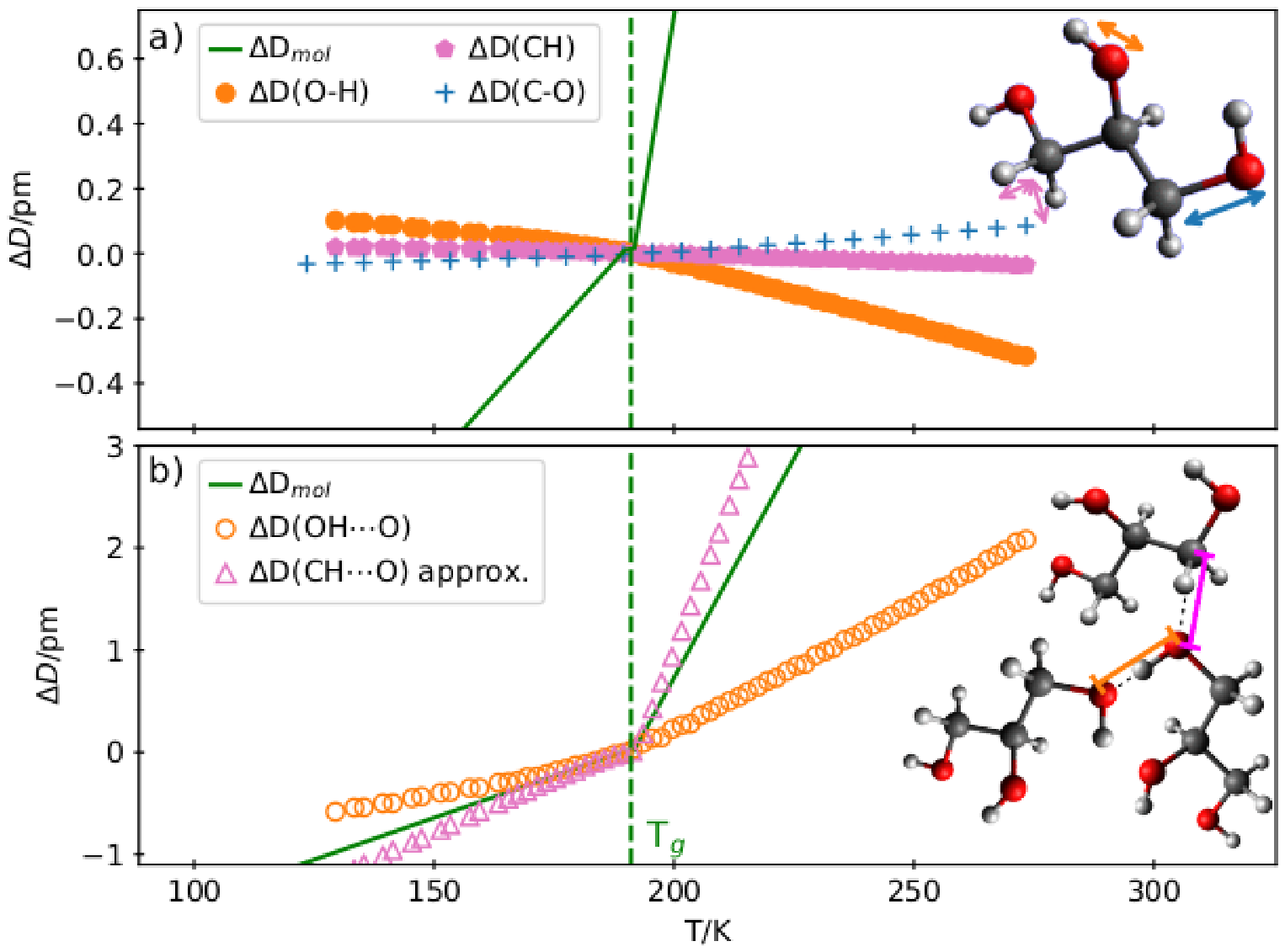}
\includegraphics[width=8cm]{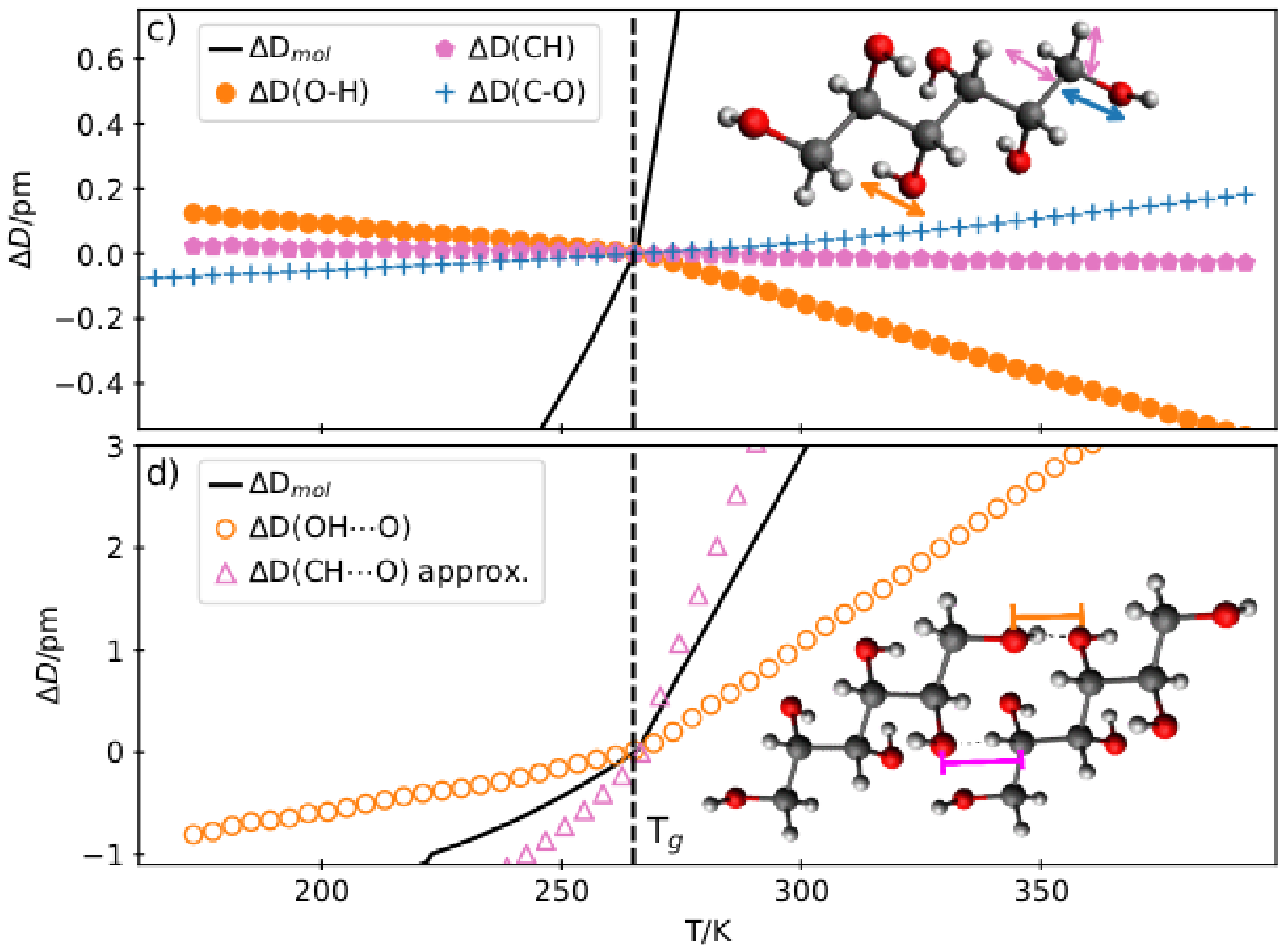}
\caption{Absolute expansion of a) \textit{intra}- and b) \textit{inter}molecular distances, bond and hydrogen bridge lengths in glycerol, and c) and d) in sorbitol, respectively. $\Delta D_{mol}$ is calculated from the mass densities \cite{footnote_density} taken from \cite{blazhnov2004,naoki1993}. $T_g$ is indicated by a dashed vertical line. The arrows in the sketched molecules indicate the respective stretching vibrations of the bonds and lengths of the hydrogen bridges (same color code as the symbols). The uncertainty originating from the experimental error of the IR measurements is smaller than the symbol size.} 
\label{fig3}
\end{figure}

For the OH bond as well as the respective OH$\cdots$O bridge however, there exists a vast record of experimental work on lengths, among them also correlations to IR absorption frequencies. Studies of crystalline H-bonding materials combined X-ray and neutron diffraction with IR spectroscopy to deduce the relation between the \textit{inter}molecular OH$\cdots$O distance $D_{OH\cdots O}$ and the peak position $\bar\nu$(OH) (i.e. the respective \textit{intra}molecular OH vibration) \cite{Libowitzky1999,steiner2002}. The found empirical relation \cite{Libowitzky1999} to extract D(OH$\cdots$O) (Fig.\,\ref{fig2}a) is given by:
 \begin{equation}
 D_{OH\cdots O}(\bar \nu) =13.21 \log \left( \frac{304\cdot10^{9}}{3592-\bar \nu\cdot cm} \right) pm
 \label{ohofunction}
 \end{equation}
Furthermore, these structural investigations established a relation between $D_{OH\cdots O}$ and $D_{OH}$ \cite{steiner2002} which is used to obtain also the temperature dependence of the latter. The thermal expansion of both $D_{OH\cdots O}$ and $D_{OH}$ exhibits a characteristic kink at $T_g$ (Fig.\,\ref{fig3}a \& b), like the underlying peak frequency of $\nu$(OH). Changes in density (indicating $T_g$) originate from \textit{inter}- and \textit{intra}molecular expansion; however, their respective contribution is yet unknown. For quantitative evaluation, we use the cubic root of the specific volume obtained from the mass density of glycerol $\rho(T)$ \cite{blazhnov2004} and its molar mass $M$ as reference. The resulting length $D_{mol}=\sqrt[\leftroot{-1}\uproot{2}\scriptstyle 3]{M/(\rho N_A)}$ (where $N_A$ is Avogardo's number) resembles the average distance between the centers of adjacent molecules (i.e. a 1-dimensional equivalent to specific volume).

While vibrational spectroscopy provides no direct access to bond lengths (i.e. structural methods are required to establish a correlation), its extremely high resolution is excellent to trace changes \cite{Deng1999}.
Consequently, in the following we focus on these (absolute) changes using the respective value at $T_g$ as reference ($\Delta D(T)=D(T)-D(T_g)$). A comparison of $\Delta D_{mol}$ with $\Delta D_{OH}$, $\Delta D_{CH}$, and $\Delta D_{CO}$ reveals that all these \textit{intra}molecular lengths exhibit a much weaker thermal expansion (about a factor of 20 for $\Delta D_{OH}$ and more than a factor of 100 for $\Delta D_{CH}$ and $\Delta D_{CO}$), and thus are negligible in the densification of the material (Fig.\,\ref{fig3}a). Consequently, the expansion of \textit{inter}molecular bridges must dominate densification. However, the change in $\Delta D_{OH\cdots O}$ is by about a factor of 3 smaller than in $\Delta D_{mol}$. Considering the chemical structure of glycerol, the only other \textit{inter}molecular interactions are of van der Waals type and weak CH$\cdots$O H-bridges \cite{Steiner1995}. The red-shift of $\nu_{as}$(CH$_2$) with decreasing temperature indicates the latter. It is known that this type of bridge has a length of $\sim350\,pm$ \cite{Steiner1995}, considerably longer than $D_{OH\cdots O}$ ($\sim280\,pm$). Attempts to establish a relation between $\nu_{as}$(CH$_2$) and $D_{CH\cdots O}$, based on studies of crystalline materials \cite{Steiner1995,Braga1995}, in order to extract a reliable temperature dependence fail because of too much scattering of the data. Additionally, due to the low polarity of the CH bond, its length is rather insensitive to the presence of H-bridge acceptors (in weak H-bridges longer than $300\,pm$, the distance between hydrogen and the H-bridge acceptor varies strongly, i.e. several tens of $pm$, while the change in donor bond length is $<1\,pm$ \cite{Steiner1998}, for -CH$\cdots$O even $<0.1\,pm$ \cite{Sosa2002,Braga1995}).

The overall thermal expansion $\Delta D_{mol}(T)$ is an average of all \textit{intra}- and \textit{inter}molecular expansivities. Since the former are negligible compared to the latter, we approximate $\Delta D_{mol}(T)$ as composition of \textit{inter}molecular expansions only:
\begin{equation}
\Delta D_{mol}(T) = \phi_{OH} \Delta D_{OH\cdots O}(T) + \phi_{CH} \Delta D_{CH\cdots O}(T)
\label{eq_rho_OHO_CHO} 
\end{equation}
Here, $\phi_{OH}$ and $\phi_{CH}$ are weighing factors representing the fraction of OH and CH donors per molecule, respectively (e.g. for glycerol $\phi_{OH} = 3/8$ and $\phi_{CH}$ = 5/8). Insertion of $\Delta D_{mol}$ obtained from density and $\Delta D_{OH\cdots O}$ calculated from eq.\,(\ref{ohofunction}) yields an approximation for the thermal expansion $\Delta D_{CH\cdots O}(T)$ (Fig.\,\ref{fig3}b) which is by about a factor of 6 larger than $\Delta D_{OH\cdots O}(T)$. Despite the lack of direct experimental data it is conceivable: due to their increased length and low polarity the CH$\cdots$O bridges are more susceptible to temperature change than the OH$\cdots$H bridges. Also, studies of semi-crystalline poly(3-hydroxybutyrate) reported thermally induced changes of inter-chain distance in the crystallites, which is governed by CH$\cdots$O contacts (of methyl groups and ketones, i.e. different chemical structure), of $\sim10\,pm$ accompanied by shifts in $\nu$(CH$_2$) of $\sim3\,cm^{-1}$ \cite{Sato2006} (the expansion of $\Delta D_{CH\cdots O}$ estimated from eq.\,(\ref{eq_rho_OHO_CHO}) is $\sim3\,pm/cm^{-1}$).

In sorbitol, the thermal expansion of the different \textit{intra}- and \textit{inter}molecular lengths is almost identical to those in glycerol (Fig.\,\ref{fig3}). The only deviation is a slightly larger rate of $\Delta D_{OH\cdots O}$ in sorbitol, probably due to the elevated temperature. This striking similarity indicates that, despite increased molecular weight and number of conformations, density is controlled by the same mechanism. Although we have no extended density data for xylitol and threitol available, the similarity of their IR band evolution suggest an analogous picture. Hence, the view \cite{footnote_typeA-B-glass,*Lunkenheimer:2002a,*Geirhos:2018,*Kudlik:1999,*Loidl2018} is corroborated that glassy densification in these polyalcohols follows the same physical mechanisms \cite{Doess:2001,Hensel:2004,Pronin:2010}.

\begin{figure}
\centering
\includegraphics[width=8cm]{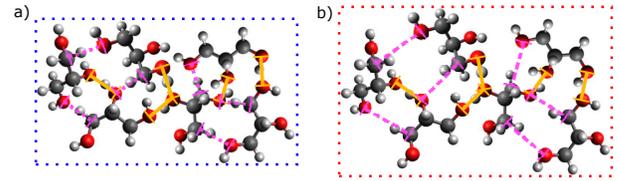}
\caption{Sketch of glycerol molecules to exemplify the transition from a) higher to b) lower density (dotted boxes visualize the specific volume of the same six molecules) at minimal expansion of strong OH$\cdots$O H-bridges (orange solid lines). This is realized by expansion of weak CH$\cdots$O H-bridges (magenta dashed lines) and possibly conformational changes.} 
\label{fig4}
\end{figure}

Despite the fact that the biggest thermal expansion of these polyalcohols is found in the CH$\cdots$O bridges, $\bar\nu_{CH}(T)$ does not exhibit a characteristic feature at $T_g$ - in contrast to the pronounced kink at $T_g$ in $\bar\nu_{OH}(T)$. Consequently, the picture is nurtured that the OH$\cdots$O bridges govern dynamics since these pose the largest energetic barriers to structural relaxation while their impact on overall thermal expansion is small (evidenced by the reduced temperature dependence of density with increasing number of H-bridging groups \cite{Gromnitskaya2019}). That is because (weak) CH$\cdots$O bridges enable considerable changes in average \textit{inter}molecular distance and thus density (Fig.\,\ref{fig4}) while the network topology is stabilized by mildly expanding (strong) OH$\cdots$O bridges.

That may illuminate a yet unresolved observation reported in the frame of the density scaling concept \cite{Floudas:2010,grzybowski2018universality}. In this empirical approach, the structural relaxation time is considered to scale with the term $\rho^\gamma/T$ with the exponent $\gamma$ \cite{Floudas:2010,Ingebrigtsen2012}. While many van der Waals liquids obey this scaling \cite{Paluch2014,Hensel:2004}, particularly associating liquids (and some polymers) defy it \cite{Floudas:2010,Pawlus2011,Paluch2014,Hensel:2004}. This defiance is caused by the difference in thermal expansion of strong and weak \textit{inter}molecular bridges; their selective impact on density and structural relaxation loosens the relation between the latter two properties which in turn disrupts the scaling. From the interpretation that $\gamma$ is connected to the (uni-form) \textit{inter}molecular potential \cite{Floudas:2010,Pedersen2008}, one can infer that the reason is the negligence of different (or inhomogeneous) \textit{inter}molecular interactions. This corroborates previous views, that glycerol in particular and possibly associating liquids in general do not comply with common theories of glassy densification \cite{Niss:2018}. Reports of separated rotational and translational dynamics in glycerol attributed to the H-bridge network fortify this \cite{Meier:2012,Flaemig:2019}. Since density scaling also fails in some polymers \cite{Paluch2003}, our results likely reach beyond associating liquids.

In summary, the presented study evidences inhomogeneities of the thermal expansion on both \textit{intra}- (Fig.\,\ref{fig3}a \& c) and \textit{inter}molecular (Fig.\,\ref{fig3}b \& d) scale in the course of glassy densification. In turn, two fundamental presumptions which are inherent to most theoretical models of molecular liquids and their glass transition are checked experimentally (for short polyalcohols): firstly, it is confirmed that \textit{intra}molecular expansion is negligibly small compared to \textit{inter}molecular expansion. Secondly, the assumption of isotropic \textit{inter}molecular interactions is an oversimplification in these systems. In contrast to the extreme cases of glass formation where the inhomogeneity of thermal expansion is negligible, i.e. inorganic glasses (like silica) characterized by extremely strong covalent bonds on the one hand and van der Waals glasses dominated by rather uniform weak interactions on the other hand, many molecular glasses contain both strong and weak interactions simultaneously. Particularly, the different impact of strong and weak bridges on structural relaxation and on density loosens the relation between the latter two properties which likely causes the open questions in the description of associating liquids \cite{grzybowski2018universality}. All this should motivate the consideration of distinct atomistic interactions in order to gain a molecular understanding of the dynamic glass transition in organic matter. To what extent these findings apply beyond associating liquids to complex glass forming systems in general, remains a challenge for future experimental and theoretical work.

\section{Acknowledgement}
Financial support by the German Science Foundation (DFG) within the collaborative research center SFB TRR 102, sub-projects B08 and B15, respectively, is highly appreciated.

\bibliographystyle{achemso} 
\bibliography{IR-bib}

\begin{mcitethebibliography}{82}
\bibitem[foo()]{footnote_IR}
Generally, a shift in band position of a stretching vibration can have
  different causes; however, gradual shifts on heating/cooling reflect changes
  in respective bond length.\relax
\mciteBstWouldAddEndPunctfalse
\mciteSetBstMidEndSepPunct{\mcitedefaultmidpunct}
{}{\mcitedefaultseppunct}\relax
\EndOfBibitem
\bibitem[foo()]{footnote_neutrons}
We note that the following neutron scattering study, which employed fits of
  atom correlation models, claims that glycerol does not form CH$\cdots$O
  bridges; however, this conclusion is indirect and incompatible to the
  red-shift observed in the $\nu_{as}$(CH$_2$) band.\relax
\mciteBstWouldAddEndPunctfalse
\mciteSetBstMidEndSepPunct{\mcitedefaultmidpunct}
{}{\mcitedefaultseppunct}\relax
\EndOfBibitem
\bibitem[foo()]{footnote_typeA-B-glass}
Particularly, our findings oppose earlier views that, based on their dielectric
  response, glycerol and sorbitol might represent two different types of glass
  formers (type A and type B glasses, respectively).\relax
\mciteBstWouldAddEndPunctfalse
\mciteSetBstMidEndSepPunct{\mcitedefaultmidpunct}
{}{\mcitedefaultseppunct}\relax
\EndOfBibitem
\end{mcitethebibliography}

\end{document}